\documentclass{eptcs}
\providecommand{\event}{PrePost 2017} 
\usepackage{underscore}           
\usepackage{tikz}
\usepackage{graphics}
\usepackage{xspace}
\usepackage[T1]{fontenc}
\usepackage{lmodern}
\usepackage{cleveref}

\newcommand{\wrt}{\textsl{wrt.}\xspace}
\newcommand{\ie}{\textsl{ie.,}\xspace}
\newcommand{\eg}{\textsl{eg.,}\xspace}
\newcommand{\aka}{\textsl{aka.}\xspace}
\newcommand{\etc}{\textsl{etc.}\xspace}
\def\cmark{\tikz\fill[scale=0.4](0,.35) -- (.25,0) -- (1,.7) -- (.25,.15) -- cycle;} 

\begin{document}

\title{A Survey of Runtime Monitoring Instrumentation Techniques\thanks{The work presented in this paper was partly supported by the project ``TheoFoMon: Theoretical Foundations for Monitorability'' (grant number: 163406-051) of the Icelandic Research Fund.}}
\author{Ian Cassar\thanks{The research work disclosed in this publication is partially funded by the
		ENDEAVOUR Scholarships Scheme. ``The scholarship may be part-financed by the
		European Union --- European Social Fund''}
	\institute{University of Malta, Malta}
	\institute{Reykjavik University, Iceland}
	\email{ian.cassar.10@um.edu.mt}
	\and
	Adrian Francalanza
	\institute{University of Malta, Malta}
	\email{adrian.francalanza@um.edu.mt}
	\and
	Luca Aceto
	\institute{Reykjavik University, Iceland}
	\institute{Gran Sasso Science Institute, L'Aquila, Italy}
	\email{luca@ru.is}
	\and
	Anna Ing\'{o}lfsd\'{o}ttir
	\institute{Reykjavik University, Iceland}
	\email{annai@ru.is}
}
%

\def\titlerunning{A Survey of Runtime Monitoring Techniques}
\def\authorrunning{I. Cassar et al.}
	
	\maketitle
	\begin{abstract}
	Runtime Monitoring is a lightweight and dynamic verification technique that involves observing the internal operations of a software system and/or its interactions with other external entities, with the aim of determining whether the system \emph{satisfies or violates a correctness specification}. Compilation techniques employed in Runtime Monitoring tools allow monitors to be automatically derived from high-level correctness specifications (\aka \emph{properties}). This allows the same property to be converted into \emph{different types of monitors}, which may apply different instrumentation techniques for checking whether the property was satisfied or not. In this paper we compare and contrast the various types of monitoring methodologies found in the current literature, and classify them into a spectrum of monitoring instrumentation techniques, ranging from completely asynchronous monitoring on the one end to completely synchronous monitoring on the other.
	\end{abstract}

\section{Introduction}
%
Formally ensuring the correctness of modern-day concurrent systems is an arduous task, mainly because exhaustive methods such as model-checking quickly run into state-explosion problems; this is typically caused by the multiple thread interleavings of the system being analysed, and the range of data the system can input and react to.  Runtime Monitoring \cite{Leu:RV:Overv} is an appealing compromise towards ensuring correctness, as it circumvents such scalability issues by verifying only the \emph{current} system execution. Runtime Monitoring thus employs techniques for observing the internal operations of a software system and/or its interactions with other external entities with the aim of determining whether the system \emph{satisfies or violates a correctness specification}. Most runtime monitoring frameworks \cite{java-mac,Havelund2004,Chen2005,Cassar2017Betty,FraSey14,elarva:2012,diana04,rover} employ compilation techniques that synthesise \emph{monitors} from high-level correctness specifications (\aka \emph{properties}), expressed in terms of a \emph{formal logic} --- the synthesised monitors then execute in tandem with the system. Automatic monitor synthesis permits for the same property to be converted into a monitor which may apply \emph{different instrumentation techniques} for observing the system. This allows the designated monitor to determine whether the property was satisfied or violated. 

In this paper we therefore examine the different types of monitoring instrumentation techniques found in the current literature. In \Cref{sec:off-on} we establish the terminology that we will be using throughout this paper and provide the necessary preliminary material for better understanding the monitoring instrumentation approaches identified in our survey. Particularly, we disambiguate between online and offline monitoring and we go into detail \wrt the former, by defining what we mean when referring to synchronous and asynchronous monitoring. 
In \Cref{sec:spectrum} we then present a \emph{spectrum of online monitoring instrumentation techniques} \wrt concurrent (component-based) systems, which is based on the different definitions that one can find in the current literature. This spectrum classifies the identified monitoring instrumentation techniques by the level of coupling and control that they posses over the system's components. In \Cref{sec:tools} we then analyse the monitoring approaches employed by state-of-the-art runtime monitoring tools, \wrt our classification of monitoring approaches. Finally, in \Cref{sec:conc} we conclude with a summary of our contributions.

\section{Preliminaries}\label{sec:off-on}
In this section we provide the necessary background material for understanding our survey. Particularly, we look into offline monitoring as this is often confused with asynchronous online monitoring, while we also discuss online monitoring in terms of the different definitions that one can find in the current literature.

\subsection{Offline and Online monitoring}
Runtime monitoring can immediately be divided into two main classes, namely \emph{Online} and \emph{Offline monitoring}.
%
In offline monitoring \cite{Leu:RV:Overv,chen-rosu-2007-oopsla,Exago,MeredithR10} the system is not directly monitored while it is executing. Instead, the relevant system events are recorded as an execution trace inside a data store. Once the monitored system terminates (or whenever a satisfactory number of events have been recorded), the collected execution trace is forwarded to the offline monitor. The offline monitor, which is entirely independent from the system, then proceeds by inspecting the system events recorded in the trace. Provided that the trace provides enough information, the monitor is able to deduce whether the correctness property was satisfied or violated.

Offline monitors are particularly suitable for properties that can only be verified by \emph{globally analysing} the \emph{complete execution trace} that is generated once the system stops executing. In fact, some properties may require to be globally analysed using a \emph{backward traversal} \cite{MeredithR10,SyncVSAsync:Rosu:2005} of the trace, and occasionally requires going back and forth along this trace during analysis \cite{SyncVSAsync:Rosu:2005}. This monitoring technique is less intrusive compared to online monitoring, as it does not interfere with the system except for the logging of events, thus imposing little runtime overhead. This, however, comes at the cost of \emph{late detections}, since violations can only be detected once the system \emph{stops executing}, which is when the recorded trace can be inspected by the monitor. Hence, this monitoring approach is generally more suited for double checking the behaviour of systems that already possess a relatively high correctness confidence. 

Unlike offline monitoring, in an online monitoring setup \cite{Leu:RV:Overv,lola:runtime,scala}, an executing system is dynamically monitored for violations (or satisfactions) \emph{during the course of its execution}. Online monitors are therefore developed to execute alongside the system by verifying its execution in an \emph{incremental fashion} \cite{Leu:RV:Overv}. This means that the monitor must be able to receive notifications about relevant events occurring in the executing system and make a decision based on the information it has collected so far. 

As online monitors are developed to verify currently executing systems, they are also capable of making early detections. Achieving early detections is crucial when monitoring for security and critical properties, which may require immediate system reparation when violated. For instance, monitoring for the correct operation of safety critical systems such as a power plant controller would require prompt detections followed by immediate reparation, since late reaction to errors might lead to serious consequences.

Early detections are thus, often, exploited by runtime adaptation tools \cite{Cassar2015,Cassar2016IFM} to administer adaptation actions with the aim of mitigating the damage incurred by the system as a result of a property violation. The primary disadvantage of online monitoring is that it imposes an inevitable runtime overhead on the system 
given that additional monitoring code is added to the system. Since runtime overheads are an undesirable side effect, a substantial effort is generally devoted to create efficient monitors \cite{Cassar2014Foclasa,Cassar2015Fesca,diana04} that minimise this overhead.

\subsection{Varying definitions of Online Monitoring}
\label{sec:syn-asyn}
In the current literature we often encounter different definitions for online monitoring, which often encapsulate the type of instrumentation protocol employed for observing the monitored system. In this paper, we follow the definitions for \emph{asynchronous monitoring} given in \cite{cc-saferAsync,SyncVSAsync:Rosu:2005,elarva:2012, Cassar2014Foclasa,Cassar2017Betty}. These definitions state that although the synthesised monitors still execute alongside the system, they are however \emph{loosely-coupled} with the system to the extent that they barely have any control over the monitored system and may suffer from \emph{late detections}. In fact whenever a specified event is performed by the system, an event notification is asynchronously forwarded to the monitor without suspending any part of the system. The monitor is then able to verify the received event notifications at its own pace, and independently from the system. It is important to note that, unlike offline monitors, asynchronous monitors execute alongside the system and analyse the system's trace of events in an incremental fashion, whereas offline monitors verify properties over a pre-recorded trace. Due to this lack of synchrony and control, asynchronous monitors impose considerably lower overheads compared to other approaches as shown in \cite{Cassar2014Foclasa}. However, most of the time an asynchronous monitor is unable to detect violations immediately and carry out effective mitigation actions \cite{Cassar2015,Cassar2016IFM}. Hence, we regard instrumentation protocols as being ideal for achieving asynchronous monitoring if they enable the monitor to observe the system without requiring (parts of) it to wait until the necessary monitoring logic ends executing.

Conversely, instrumentation techniques applied for achieving synchronous monitoring should permit the monitor to employ some synchronisation mechanism, which delays system execution until the necessary monitoring code executes. However, synchronous instrumentation protocols usually vary in two aspects, namely, in the level of \emph{coupling} between the monitor and the system, and in the level of \emph{control} that the monitor has over the system. For instance, synchronous online monitoring is sometimes presented as being very tightly coupled \cite{taxonomy:Delgado:2004,lola:runtime,BauFal12} to the system and as having a very high level of control over the system, to the extent that it causes \emph{all} system components to block even when only a single component generates a monitorable event. Given that this extreme synchronous instrumentation technique uses a large amount of synchronisation, it generally imposes substantial amounts of overheads and is thus more appealing for inherently synchronous systems \cite{lola:runtime,BauFal12} such as circuits and embedded systems.  Other research work was conducted to reduce monitoring overheads by implementing a less stringent definition of synchronous online monitoring. Definitions along the lines of those in \cite{Cassar2014Foclasa,chen-rosu-2007-oopsla,java-mac,Havelund2004,Chen2005} allow the monitors to only block the concurrent component that generates the specified event, without influencing the execution of the other components; other definitions, such as those in \cite{cc-saferAsync,Cassar2014Foclasa}, invoke synchronisation at specific points to allow the lagging monitor to keep up with the system. In the definitions given in \cite{SyncVSAsync:Rosu:2005,Cassar2014Foclasa}, synchronisation is employed only as a way to ensure the timely detection of security-critical properties, \ie in cases where late detections are infeasible.

\section{A Spectrum of Online Monitoring Instrumentation approaches}
\label{sec:spectrum}
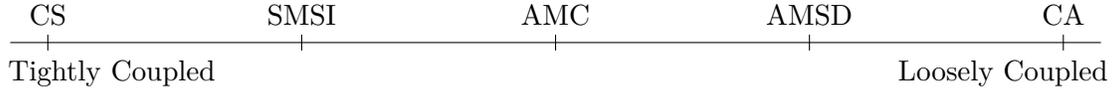
\begin{figure*}
	\centering
	\begin{tikzpicture}
	\draw (-5,0) -- (9.5,0);
	
	\foreach \x in {-4.5,-1.125,2.25,5.625,9}
	\draw (\x cm,3pt) -- (\x cm,-3pt);
	
	\draw (-3.65,0) node[below=3pt] {Tightly Coupled};	
	\draw (-4.5,0) node[above=3pt] {CS}; 
	\draw (-1.125,0) node[above=3pt] {SMSI}; 
	\draw (2.25,0) node[above=3pt] {AMC}; 
	\draw (5.625,0) node[above=3pt] {AMSD}; 
	\draw (9,0) node[above=3pt] {CA}; 
	\draw (8.2,0) node[below=3pt] {Loosely Coupled};	
	\end{tikzpicture}\vspace{-2mm}
	\caption{The Online monitoring Spectrum.\label{fig:pic}}
	\vspace{-2mm}
\end{figure*}

Based on the range of definitions surveyed in \Cref{sec:off-on}, we devise a \emph{spectrum} of online monitoring approaches, which we explain in the setting of \emph{component-based systems}, whereby components represent concurrent entities (\eg node, thread, actor, \etc) that can be in either of two states, namely \emph{blocked} or \emph{running}. As depicted in \Cref{fig:pic}, this spectrum ranges from a tightly coupled \emph{completely-synchronous (CS)} monitoring instrumentation approach on the one end, to a loosely-coupled \emph{completely-asynchronous (CA)} monitoring approach on the other end. 

The spectrum also presents the currently known online monitoring approaches that lie in between these two extremes; such approaches includes: \emph{Synchronous Monitoring with Synchronous Instrumentation (SMSI)}, \emph{Asynchronous Monitoring with Checkpoints (AMC)} and \emph{Asynchronous Monitoring with Synchronous Detections (AMSD)}. These intermediate approaches provide a trade-off between the level of coupling and control that the monitor has upon the system. It is generally assumed that monitoring approaches that are closer to the synchronous end of the spectrum tend to posses a higher level of control over the system, which comes at the expense of higher overheads. Conversely, approaches that are closer to the opposite end are generally assumed to be more efficient, yet lack a high degree of control over the monitored system. This assumption was investigated and confirmed in \cite{Cassar2014Foclasa,Zhang2016SyncVsAsync}, after conducting a series of empirical experiments using various levels of synchronisations in the employed monitoring mechanisms.

In the following subsections, we explain each monitoring instrumentation technique defined within our spectrum \wrt an example property that assumes a simple transaction system that must satisfy the following invariant:
\begin{center}
	\vspace{-2mm}``\emph{A user cannot make a transaction before a login.}''	
\end{center}
This property is thus violated whenever a component generates a transaction event \textsf{trans} \emph{before} executing a \textsf{login} event.

\subsection{Completely-synchronous Monitoring (CS)}
As depicted in \Cref{fig:pic}, \emph{Completely-synchronous Monitoring} \cite{taxonomy:Delgado:2004,lola:runtime,BauFal12} refers to the synchronous extreme of the spectrum. As shown in \Cref{CS} below, this implies that the system and the monitors are so \emph{tightly coupled} together to the extent that whenever an event occurs in \emph{one} component of the system, such as $C1$, the \emph{entire} system 
execution is \emph{interrupted} (as shown by \textsl{(1)}). The system components remain blocked until the necessary monitoring checks are performed, in which case the monitor unblocks the components (as shown by \textsl{(2)}). 

\begin{figure}[h]
	\centering
	\includegraphics[width=1\textwidth]{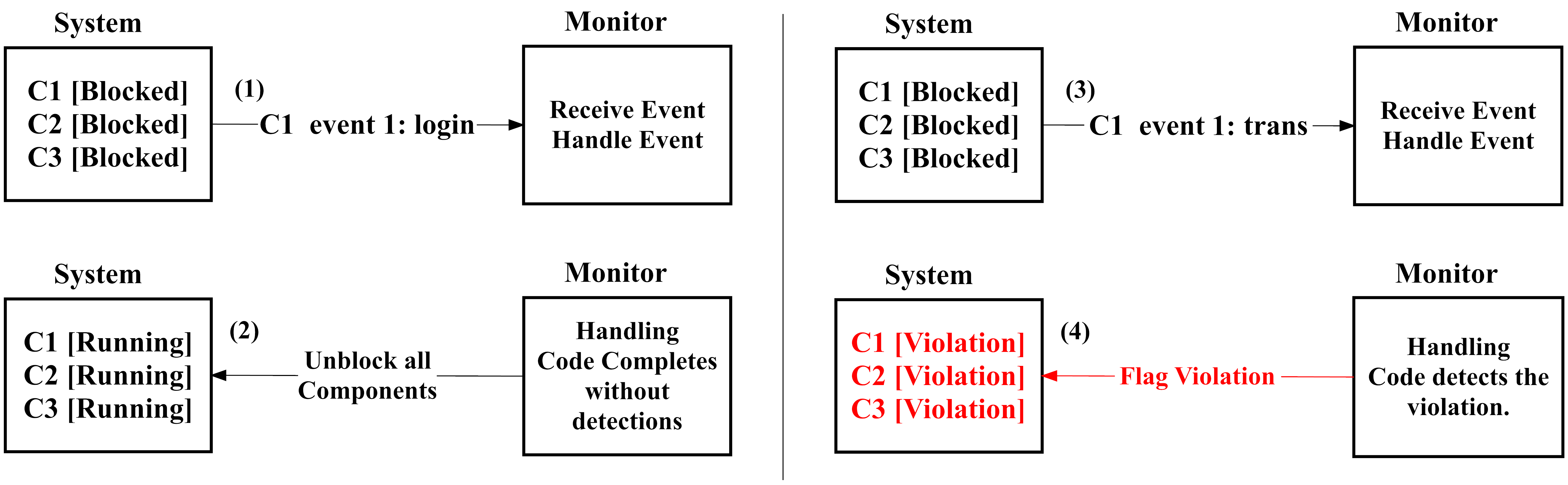}
	\caption{Completely-synchronous Monitoring (CS)}
	\label{CS}
	\vspace{-1mm}
\end{figure}

As depicted by \textsl{(3)} and \textsl{(4)}, completely-synchronous monitoring allows for our property to be immediately detected since all the system's components (including the offending component $C2$) are blocked at the time when the violation is detected. Since CS monitoring provides timely detections, it can therefore be used to effectively mitigate incorrect behaviour. This instrumentation approach is, however, highly intrusive as it introduces an unnecessarily high level of synchronisation that might lead to an infeasible performance degradation on the monitored system. Completely-synchronous monitoring is generally applied to traditional monolithic systems and also to small synchronous systems \cite{lola:runtime,BauFal12} such as embedded systems and circuits.

\subsection{Synchronous Monitoring with Synchronous Instrumentation (SMSI)}
\emph{Synchronous Monitoring with Synchronous Instrumentation} \cite{java-mac,Havelund2004,Chen2005,chen-rosu-2007-oopsla,Cassar2014Foclasa} is a monitoring approach that is close to a completely-synchronous approach, yet is less intrusive. In fact as shown in \Cref{SMSI} below, this approach assumes that whenever a \emph{single component} of the monitored system, such as $C1$, 
executes a specified event, then \emph{only the execution of this component} is interrupted (as shown by \textsl{(1)}). This component remains blocked until monitoring completes, in which case the component is reset to a running state (as shown by \textsl{(2)}). 

\begin{figure}[h]
	\centering
	\includegraphics[width=1\textwidth]{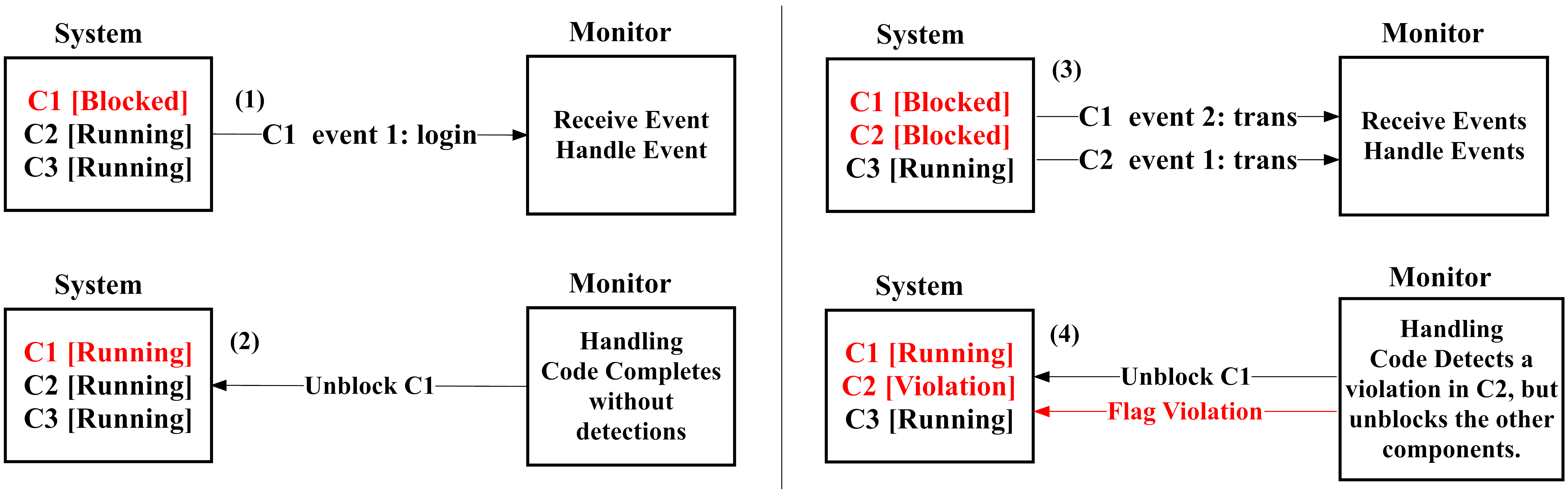}
	\caption{Synchronous Monitoring with Synchronous Instrumentation (SMSI)}
	\label{SMSI}	\vspace{-1mm}
\end{figure}

As illustrated by \textsl{(3)} and \textsl{(4)}, a violation is only detected for component $C2$ since this executes a \textsf{trans} event without a preceding \textsf{login} event --- once again this violation is detected in a timely fashion since $C2$ was blocked by the monitor upon generating the offending event. $C1$, however, was allowed to proceed with its execution given that it executed a correct event sequence, \ie a \textsf{login} followed by a \textsf{trans} event.

Synchronous instrumentation monitoring hence reduces the level of overall synchronisation by relinquishing control over the other components, thereby allowing them to keep executing whenever a certain component generates an event. However, this monitoring technique still possesses a relatively high level of control over the system, as it still needs to \emph{synchronously inspect each and every event}, even though it only interrupts the component, that generated the event, until monitoring completes. 

\subsection{Asynchronous Monitoring with Checkpoints (AMC)}
\begin{figure}[h]
	\centering
	\includegraphics[width=0.55\textwidth]{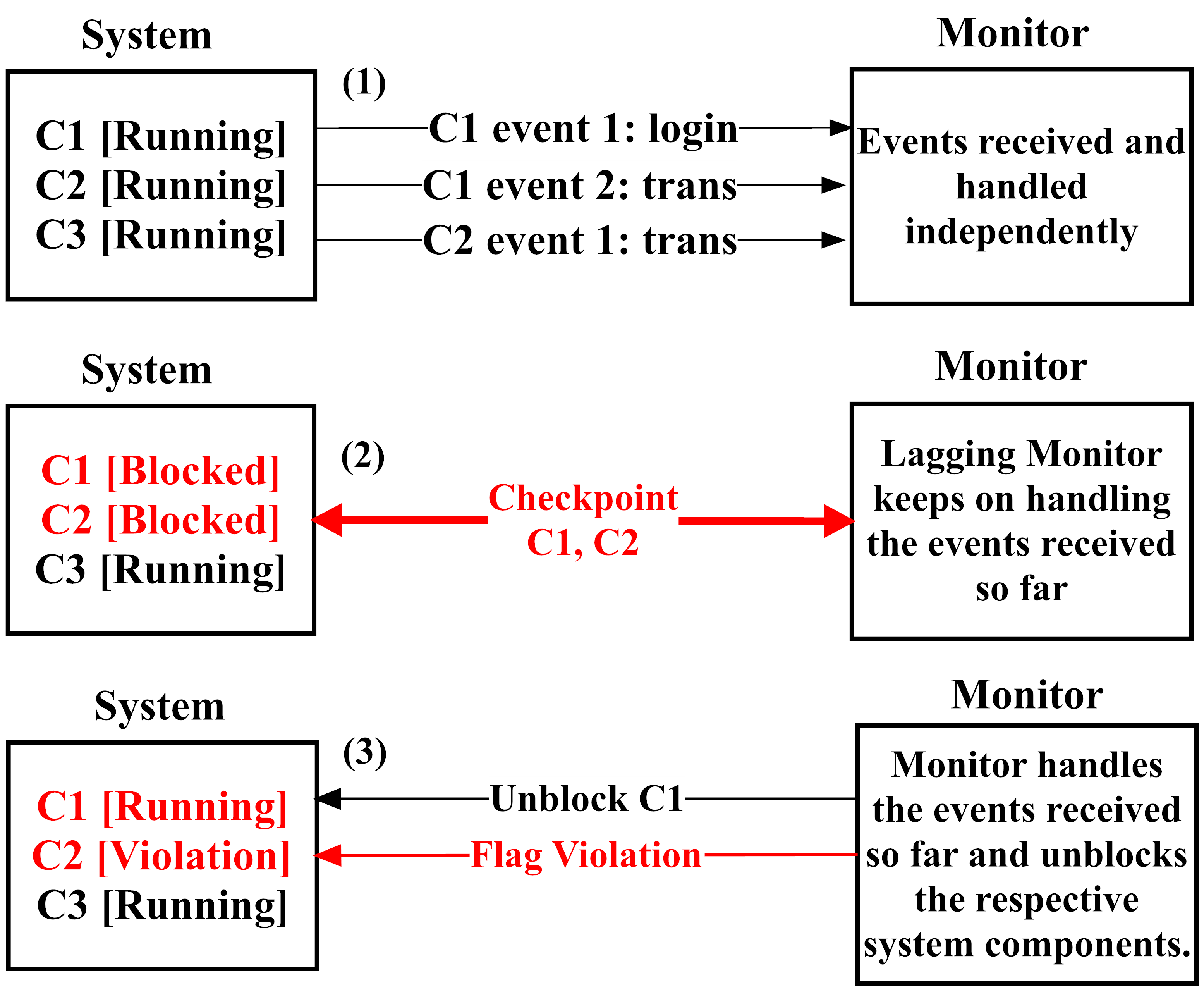}
	\caption{Asynchronous Monitoring with Checkpoints (AMC)}
	\label{AMC}	\vspace{-1mm}
\end{figure}

This approach \cite{cc-saferAsync,taxonomy:Delgado:2004} allows for an asynchronous decoupling between the system (or parts of it) and monitor executions, while also providing the user with the ability to specify checkpoints where the decoupled system and monitor executions should synchronise. We conjecture that, for component-based systems, checkpoints may also be associated with specific system components. For instance, as shown by \textsl{(1)} and \textsl{(2)} in \Cref{AMC}, a specifier may use a checkpoint to specify that certain system components, such as $C1$ and $C2$, should temporarily block their execution and synchronise with the monitor. Such synchronisation can be done during periods in which the system is not required to be responsive, and therefore permits waiting for the monitor to catch up with the system execution. 

As shown by \textsl{(3)}, once the monitor synchronises with the system, it unblocks component $C1$ and allows its execution to proceed, since it executes a \textsf{trans} action after a \textsf{login}. The monitor, however, flags a violation for $C2$ and opts not to unblock the assigned checkpoint thereby allowing the violation to be timely detected. An appealing feature of this monitoring approach is that it enables the user to manually determine the level of synchrony and control that the monitor can possess over the system. The user may also use this mechanism to specify that the system should stop and wait for the monitor's analysis to complete when certain \emph{safety-critical events} occur during execution \cite{cc-saferAsync}. This ensures that safety-critical violations are timely detected, thus permitting the monitor to effectively react to the violation and possibly invoke mitigating actions. To further reduce the level of monitor intrusion, 
the user may specify that the system should keep on executing whenever it generates non-safety critical events. 

\subsection{Asynchronous Monitoring with Synchronous Detection (AMSD)}
\emph{Asynchronous Monitoring with synchronous detection} \cite{Cassar2014Foclasa,SyncVSAsync:Rosu:2005} is yet another monitoring approach that lies close to completely-asynchronous monitoring in our spectrum. As completely-asynchronous monitors sacrifice timely detections for efficiency, this intermediate approach attempts to introduce synchronous (timely) detections into asynchronous monitoring. The system may therefore send event notifications without blocking in the same way as in completely-asynchronous monitoring. The system is, however, required to synchronously monitor for critical events that may directly lead to a property violation. This approach uses minimal synchronisation to ensure synchronous (timely) detections, since synchronisation is employed \emph{only} for system events that might \emph{directly contribute to a violation}. 

\begin{figure}[h]
	\centering
	\includegraphics[width=0.6\textwidth]{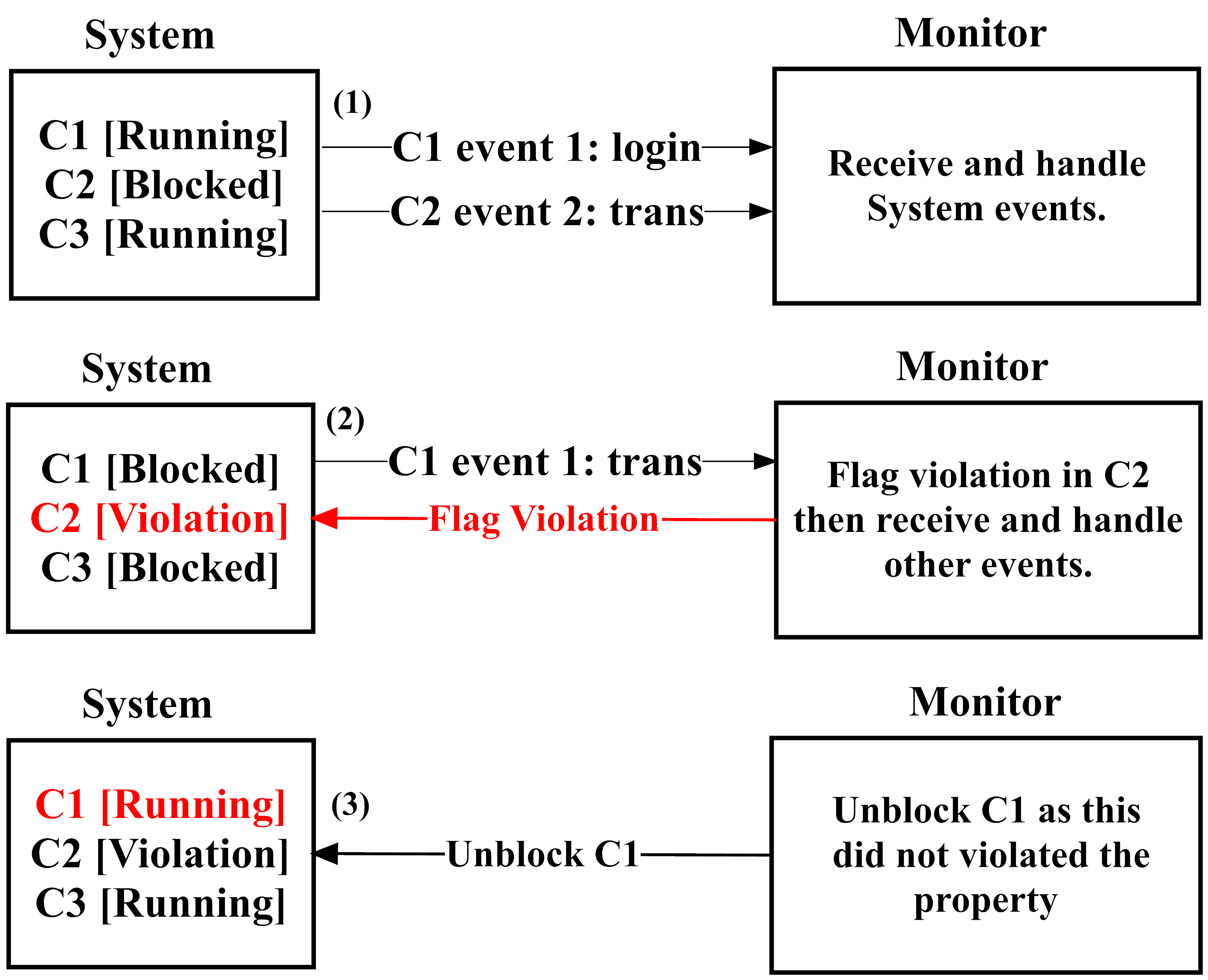}
	\caption{Asynchronous Monitoring with Synchronous Detection (AMSD)}
	\label{AMSD}	
\end{figure}

As our property is only violated whenever a transaction event (\textsf{trans}) is observed without a preceding \textsf{login} event, to detect violations of our property in a timely fashion, it suffices that the monitor synchronises only with a system component whenever the component performs a transaction event. As shown by \textsl{(1)} in \Cref{AMSD}, only component $C2$ is blocked, because it produced a transaction event that may lead to a property violation. By contrast $C1$ kept on executing since the \textsf{login} event it produced does not directly lead to a detection. In \textsl{(2)} we see that the monitor managed to detect that $C2$ violated our property, as it did not execute a \textsf{login} action prior to performing the transaction. The detection was thus achieved in a timely manner since $C2$ was not allowed (in \textsl{(1)}) to execute any further after committing the transaction event that led to the violation. Furthermore in \textsl{(2)} one can notice that $C1$ blocked waiting for the monitor's verdict as a result of producing a transaction event, in which case the monitor detects that $C1$ had already performed a \textsf{login} event meaning that it did not violate our property. Hence in \textsl{(3)} the monitor allows $C1$ to proceed. 

\subsection{Completely-asynchronous Monitoring (CA)}
\begin{figure}[h]
	\centering
	\includegraphics[width=0.6\textwidth]{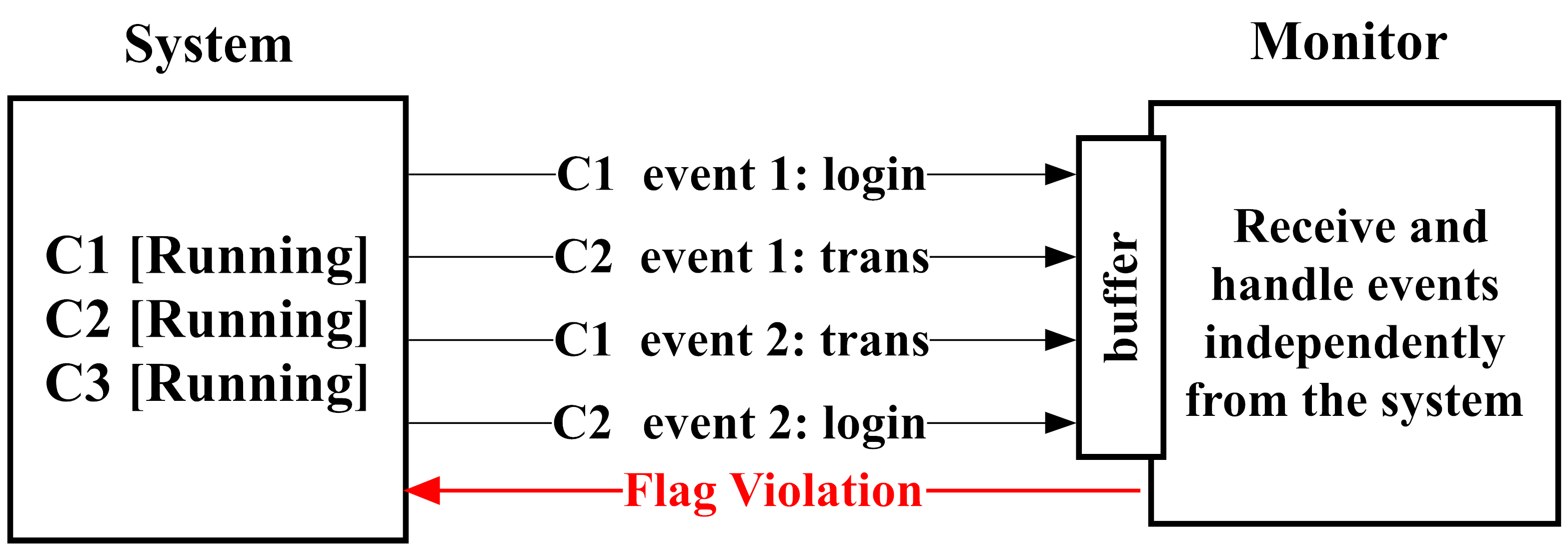}
	\caption{Completely-asynchronous Monitoring (CA)}
	\label{CA}	
\end{figure}

Finally, in \emph{Completely-asynchronous Monitoring} \cite{Cassar2017Betty,FraSey14,Cassar2014Foclasa,elarva:2012} the monitors are designed to be as \emph{loosely coupled} as possible from the system they are monitoring. In fact completely-asynchronous monitors are designed to listen for system events and handle them in the background without interfering in any way with the system execution. Hence, as shown in \Cref{CA}, in completely-asynchronous monitoring, the system is allowed to proceed immediately after placing an event notification (\eg $C1 \textsf{ login}$) in the monitor's buffer. The monitor can then independently read the event notifications from its buffer and carry out the necessary checks at its own pace. 

Although this approach imposes minimal intrusion in the monitored system, it suffers from late detection. This is shown in \Cref{CA}, since the violation performed by component $C2$ was detected after it managed to produce a subsequent \textsf{login} event. This makes it highly unlikely for an asynchronous monitor to be able to effectively mitigate the detected misbehaviour. As offline monitoring \cite{taxonomy:Delgado:2004, SyncVSAsync:Rosu:2005} is inherently asynchronous, it is sometimes classified as being an asynchronous monitoring approach. However, we conjecture that offline monitoring generally requires pre-recorded (generally complete) traces, while CA monitoring can \emph{incrementally analyse partial traces} that can be extended as the system proceeds with its execution. 

\newcommand{\mon}[1]{\mathcal{M}(#1)}
\newcommand{\lphi}[0]{\mathcal{L}(\varphi)}
\newcommand{\decentmon}[0]{\textsc{DecentMon}\xspace}
\newcommand{\larva}[0]{\textsc{Larva}\xspace}
\newcommand{\elarva}[0]{e\textsc{Larva}\xspace}
\newcommand{\exago}[0]{\textsc{Exago}\xspace}
\newcommand{\clarva}[0]{c\textsc{Larva}\xspace}
\newcommand{\lola}[0]{\textsc{Lola}\xspace}
\newcommand{\eventchaser}[0]{\textsc{EventChaser}\xspace}
\newcommand{\eagle}[0]{\textsc{Eagle}\xspace}
\newcommand{\jeagle}[0]{\textsc{JEagle}\xspace}
\newcommand{\ruler}[0]{\textsc{RuleR}\xspace}
\newcommand{\jpax}[0]{\textsc{Jpax}\xspace}
\newcommand{\mop}[0]{\textsc{Mop}\xspace}
\newcommand{\detecter}[0]{\textsf{detectEr}\xspace}
\newcommand{\detecterB}[0]{\textsf{detectEr 2.0}\xspace}
\newcommand{\adapter}[0]{\textsf{adaptEr}\xspace}

\section{Current Tool Implementations}
\label{sec:tools}

Several runtime monitoring tools have been developed, some of which implement more than one monitoring approach. In this section we review a sample of these tools and classify them according to our spectrum. In order to denote the wide variety of the tools available in the current literature, the reviewed tools were selected based on: the monitoring instrumentation approaches that they implement, the technologies and architectures that they target, and the design paradigms employed to implement the monitor. 

We start our review by illustrating a comparison of the monitoring approaches that are implemented by the reviewed runtime monitoring tools in \Cref{tbl:mon-tools}. Following this we provide a detailed discussion about the monitoring approaches employed by each of these tools.

\begin{table}[h]
	\centering\vspace{4mm}
	\begin{tabular}{rccccccc}
		\hline\\[-3mm]
		&\multicolumn{6}{c}{Online}& Offline \\[2mm]
		Tool              && CS     & SMSI   & AMC    & AMSD   & CA     &         \\
		\hline\\[-2mm]
		\decentmon        && \cmark &        &        &        &        &         \\
		\lola             && \cmark &        &        &        &        & \cmark  \\
		\jeagle           &&		& \cmark &        &        &        &         \\
		DB/Temporal-Rover &&        & \cmark &        &        & \cmark & \cmark  \\
		Java-MOP          &&        & \cmark &        &        & \cmark & \cmark  \\
		\larva toolkit    &&        & \cmark &        &        & \cmark & \cmark  \\
		\detecter toolkit &&        & \cmark & \cmark & \cmark & \cmark &         \\
		\exago            &&        &        &        &        &        & \cmark  \\[2mm]
		\hline
	\end{tabular}
	\caption{A Comparison of monitoring approaches implemented in Runtime Monitoring tools.}
	\label{tbl:mon-tools}
\end{table}
\vspace{-3mm}
\paragraph{\decentmon}
\decentmon \cite{BauFal12} is a decentralized runtime monitoring tool developed in OCaml, which allows users to specify properties for \emph{synchronous distributed systems} by using \emph{standard LTL syntax}. This relieves the LTL specifications from having to include implementation details denoting specific components where the required events should occur. 

This tool achieves a \emph{completely-synchronous monitoring (CS)} approach by initially converting a given formula into $n$ small monitors, ie, one monitor for each component in the distributed system. Each of these monitors is attached to a single component and utilizes \emph{progression} (formula rewriting) to update its current version of the LTL formula according to the events occurring in the local trace of the component it is attached to. Whenever the monitors require information about events occurring in other components, they migrate by sending their current LTL formula to the monitor which is attached to the component that can generate the required event. Furthermore, since synchronous distributed systems are governed and synchronised using a global clock, the necessary monitoring is theoretically carried out in between each clock cycle. In practice the entire distributed system must block until the distributed monitors are done migrating and performing checks. This approach also guarantees that violations are detected within a maximum bounded delay that is proportional to the number of components in the synchronous distributed system. 

This tool is more suited for monitoring circuits and embedded systems, \eg in automobiles, which are usually developed using multiple parallel components that work in synchrony with each other by using a global clock.
\vspace{-3mm}
\paragraph{\lola}
\lola \cite{lola:runtime} is another runtime monitoring tool which offers both \emph{offline} and \emph{completely-synchronous (CS)} online monitoring approaches that provide bounded resources guarantees. This tool allows a user to specify the necessary correctness properties using a mixture of past-time LTL and future-time LTL. When opting for offline monitoring the tool inspects the operations occurring in an execution of the system and dumps the collected trace to a storage device once the system runs to completion. The offline monitor is then able to randomly access different parts of the collected trace, which allows it to carry out analysis algorithms, such as \emph{Backward Tracing}, which cannot be applied in an online monitoring approach. This helps in order to make more accurate detections. 

\lola efficiently implements a completely-synchronous online monitoring approach for which the space requirement is independent of the size of the execution trace, and is linear in the size of the LTL specification. The synthesised monitor is designed to work entirely in synchrony with the system. In fact, whenever the monitor is required to perform certain runtime checks, the entire system must block waiting for the monitor to finish. Similar to \decentmon, \lola is more intended to be used for monitoring of synchronous systems including circuits and embedded systems. 
\vspace{-3mm}
\paragraph{\eagle and \jeagle}
\eagle \cite{eagle} is a language-independent, rule-based, runtime verification tool which extends $\mu$-calculus with data parametrization and was specifically designed to support future and past-time logics, interval logics, extended regular expressions, state machines and real-time and data constraints. In order to support these different logics and constraints, \eagle implements the basic logical operators and the \emph{next} and \emph{previous} temporal logic operators. By implementing these primitive logical operators, the remaining operators become a matter of syntactic sugar; in fact, \eagle encodes more complex operators in terms of these primitive operators. Although \eagle was implemented as a Java library, it does not support automatic code instrumentation for the collection of events, and in fact requires the user to manually create a projection of the actual program state. The user-defined properties are then evaluated with respect to this projected program state. 

\jeagle \cite{jeagle} is an extended version of \eagle which supports automated instrumentation and object reasoning at the expense of making the tool language specific, meaning that it can only be applied to programs written in Java. This tool implements \emph{synchronous monitoring with synchronous instrumentation (SMSI)} through code inlining. In fact, \jeagle converts the specified properties into Java code and uses aspect-oriented programming to automatically inline the necessary monitoring code in a given system. In the case of multithreaded Java programs, whenever a specified event occurs in a system thread, this code inlining mechanism would cause the Java thread which generated event, to pause its normal execution and start executing the monitor's code. The thread continues executing the system's code once the monitoring code completes.
\vspace{-3mm}
\paragraph{DB-Rover and Temporal-Rover}
The DB-Rover \cite{rover, taxonomy:Delgado:2004} allows a user to specify temporal rules using either \emph{Linear Temporal Logic (LTL)}, or an extended version of LTL called \emph{Metric Temporal Logic (MTL)} which permits the specification of lower bounds, upper bounds, and ranges for discrete-time and real-time properties. 
These properties are then automatically converted into monitoring code using the \emph{Temporal-Rover} compiler. 

Temporal-Rover does not embed the monitoring code into the system, but it adds code which connects to a standalone DB-Rover remote server that carries out the necessary validation checks. Whenever a specified system event occurs at runtime, the inserted code sends a monitoring request containing the necessary information to the remote DB-Rover server, which checks the received information against the specified temporal rules. This server supports online monitoring using both $(i)$ a \emph{synchronous instrumentation (SMSI)} approach and also $(ii)$ a \emph{completely-asynchronous (CA)} approach. 

Temporal-Rover implements $(i)$ by adding synchronous TCP sends and receives into the monitored system to allow it to communicate the necessary event information to the remote server. Whenever a system thread encounters a specified event, it invokes a synchronous TCP-send operation to supply the necessary event information to the server, which causes the system thread to wait until the monitoring server issues a synchronous receive. 

Conversely, Temporal-Rover provides $(ii)$ by adding asynchronous TCP sends and receives instead. In this way whenever a system thread encounters a specified event, it immediately sends the required information to the server and keeps on executing. Furthermore, Temporal-Rover also provides offline monitoring by capturing the information in a database and then using DB-Rover to analyse the collected trace information once the system stops. DB-Rover and Temporal-Rover aim to be utilized for business and security applications.
\vspace{-3mm}
\paragraph{Java-MOP}
Java-MOP \cite{chen-rosu-2005-tacas} is a Java-based tool which implements the monitoring-oriented programming (MOP) paradigm \cite{chen-rosu-2007-oopsla}. This programming paradigm aims to combine the specification of a system together with its implementation. Although MOP employs the same concept of runtime verification, monitoring in MOP is not used for double checking a system but is instead an integral part of the system's design. MOP therefore aims to separate development concerns, by allowing a system to be first developed without any sort of verification checks, and then be augmented with the necessary monitoring checks that are automatically synthesised from a given set of formally defined correctness properties. 

The Java-MOP tool provides support for numerous specification languages including past-time and future-time LTL, extended regular expressions and others. Similar to Temporal-Rover, Java-MOP uses a client-server approach and supports \emph{offline monitoring}, \emph{completely-asynchronous monitoring (CA)} as well as \emph{synchronous instrumentation monitoring} (SMSI) achieved through inlining of verification checks.
\vspace{-3mm}
\paragraph{The \larva toolkit}
\larva \cite{CPS09larva} is yet another Java-based runtime verification tool for object-oriented systems, which provides an automata-based specification language called \emph{Dynamic Automata with Timers and Events} (DATE). This specification language allows a user to specify both real-time and discrete-time properties in terms of automata. This tool is capable of converting the same DATE specification into either an \emph{offline monitor} or into an \emph{online monitor}, where the latter utilizes a \emph{synchronous-instrumentation (SMSI)} approach. 

The tool achieves offline monitoring by converting the specification into aspect code which extracts and logs the specified system events into a database whenever they occur while the system executes. The \larva compiler also generates an offline-monitor and a replayer program, where the latter is used to extract the collected events from the database and replay them such that the offline monitor is able to analyse the recoded trace and detect any violations.

\larva also creates online monitors by using aspect-oriented programming instrumentation to automatically insert the necessary verification checks inside the system components (objects) that generate the specified events. By using code-inlining this tool achieves online monitoring through a synchronous-instrumentation approach. In the case of multi-threaded systems, whenever an object component generates an event on one particular thread, only the thread in question is required to temporarily halt its normal operations and start executing the monitoring code. When the thread finishes executing the checks, it continues with the normal system execution from the point at which it was previously interrupted. Therefore, the other threads are not affected in any way while the injected monitoring code checks the generated event. 

\elarva \cite{elarva:2012} is an Erlang-based re-implementation of \larva for actor-oriented systems. Due to the asynchronous nature of the actor-oriented paradigm offered by Erlang, this tool explores the opposite end of the online monitoring spectrum. In fact \elarva converts DATE specifications into a \emph{completely-asynchronous monitor (CA)} without adding any form of instrumentation code inside the monitored system. Instead, it uses the tracing mechanism \cite{Armstrong07} provided by the Erlang Virtual Machine to capture instances when a system actor sends or receives a specified Erlang message. Whenever a specified event occurs in an actor, the tracing mechanism sends a message to the asynchronous monitor synthesised by \elarva. The synthesised asynchronous-monitor consists in a set of monitoring actors which block waiting to receive trace messages denoting relevant events occurring on the monitored system. Whenever the monitoring actors receive a relevant event, they unblock and carry out the necessary checks independently from the monitored system. 
\vspace{-3mm}
\paragraph{The \detecter toolkit}
\detecter \cite{Cassar2017Betty,FraSey14} is a runtime monitoring framework that converts properties, expressed using Hennessey Milner Logic with recursion, into monitors for Erlang systems. Originally, \detecter was developed to synthesise \emph{completely-asynchronous} monitors that were capable of observing Erlang systems by using the asynchronous tracing mechanism \cite{Armstrong07} provided by the Erlang Virtual Machine.

Subsequent work on the tool saw the inclusion of several optimisation techniques \cite{Cassar2015Fesca} along with an extension that allows the specifier to select between \emph{completely-asynchronous monitoring (CA)}, \emph{synchronous instrumentation monitoring (SMSI)}, \emph{asynchronous monitoring with checkpoints (AMC)} and \emph{synchronous detection monitoring (AMSD)} $-$ this extended version is known as \detecterB \cite{Cassar2014Foclasa}. To implement this variety of monitoring approaches, the tracing mechanism had to be replaced by code instrumentation that was achieved through an aspect-oriented programming framework for Erlang called \textsf{eAOP} \cite{Cassar2017Eaop}. \textsf{eAOP} allowed for instrumenting the system with a custom tracing protocol that, apart from reporting events to the monitor as asynchronous messages, is also able to force certain system components to block waiting for the monitor's feedback, thereby achieving synchrony.

CA in \detecterB is implemented by completely omitting the requirement for concurrent system components to wait for the monitor's feedback. By contrast, SMSI is achieved by forcing each system component to block for \emph{every} reported event. AMC and AMSD (referred to as \emph{hybrid} in \detecterB) are achieved through an extension to the specification language that saw the inclusion of \emph{synchronous necessities} and \emph{synchronous verdicts}. Synchronous necessities are used to force the instrumented component to wait for feedback whenever the event described in the necessity is reported to the monitor. Hence, AMC is achieved since synchronous necessities serve as checkpoints that allow the lagging monitor to catch up and synchronise with the system. On the other hand, synchronous verdicts allows for achieving AMSD, since they only force system components to synchronously report events that may lead to a violation.

The synchronisation protocol introduced in \detecterB was also used to allow for \emph{adaptation actions} to be effectively applied to specific Erlang components, in a timely manner. Adaptation actions allowed the monitor to apply rectifying actions (\eg restarting or terminating misbehaving components) in order to mitigate the effects incurred by the detected violation. This extension led to the creation of a Runtime Adaptation tool called \adapter \cite{Cassar2015,Cassar2016IFM}.
\vspace{-3mm}
\paragraph{\exago}
\exago \cite{Exago} is an Erlang-based runtime monitoring tool which implements an \emph{offline monitoring} approach, to perform data mining on event logs. Ca\exago is capable of gathering the necessary log files, using abstraction and validation functions to create abstract representations of the events that occurred in the monitored system, and evaluating \emph{complete traces} of these event representations, against a monitoring state machine that verifies whether the logged events denote a valid behaviour or an invalid one. Being an offline monitoring tool, \exago imposes low runtime overheads since at runtime it only logs events, which are then verified once the entire event trace is collected.

In order to further simplify the process of manually specifying a state machine model, \exago was also extended with the ability of analysing the collected logs and generating a state machine representation which models the analysed behaviour.
\section{Conclusion} \label{sec:conc}
We have introduced a \emph{spectrum of online monitoring approaches}, and explained each approach in the setting of component based systems. The spectrum consists in \emph{completely-synchronous monitoring} on one end, which achieves an extremely high level of control over the system that could potentially jeopardise monitoring efficiency, and \emph{completely-asynchronous monitoring} on the opposite end, which provides efficient monitors with very low intrusion over the monitored system, by sacrificing the level of control over the monitored system, including timely detections. We have also identified a number of intermediate approaches that lie in between these two online monitoring extremes in our spectrum. We also conjecture that there may be other intermediate approaches apart from the ones that we identified which still need to be explored. 

The terminology established in the identified spectrum advocate for a better understanding of the design space that one needs to explore when developing a software monitoring or a runtime verification tool. Hence, our spectrum also provides guidance for future development of monitoring tools. Finally, our spectrum also helps to facilitate the comparison between monitoring instrumentation approaches. In fact, this also enabled us to analyse the monitoring instrumentation approaches, which have been implemented by a number of runtime monitoring tools, \wrt our classification.

\bibliographystyle{eptcs}
\bibliography{references}

\end{document}